# Supercritical-Xenon-Filled Hollow-Core Photonic Bandgap Fiber as a Raman-Free, Dispersion-Controllable Nonlinear Optical Medium

K. E. Lynch-Klarup,[1,*] E. D. Mondloch,[1] M. G. Raymer,[1] D. Arrestier,[2] F. Gerome,[2] and F. Benabid[2]

[1]*Oregon Center for Optics and Department of Physics, University of Oregon, Eugene, OR 97403, USA*
[2] *GPPMM group, Xlim Research Institute, CNRS, Université de Limoges, France*
*\*Corresponding author: lynchkla@uoregon.edu*

We propose supercritical xenon in a hollow-core photonic bandgap fiber as a highly nonlinear medium, and demonstrate 200 nm control over the guidance window of the fiber as the xenon goes through its supercritical phase transition. The large optical polarizability and monoatomic nature of xenon are predicted to allow large optical nonlinearity, on the same order as that of fused silica, while maintaining greatly reduced Raman scattering, offering benefits for nonlinear and quantum optics.

Solid-core photonic crystal fibers (PCF) are invaluable for nonlinear optics because of their designable guidance windows and dispersion properties over large wavelength ranges [1]. A major drawback of these fibers, as well as standard silica fibers, for nonlinear optics and quantum optics is the spontaneous Raman scattering associated with vibrational transitions in the silica, deleterious in experiments on single-photon generation, parametric amplification, and squeezing in soliton propagation [2]. Because Raman scattering does not exist in atomic noble gases, the ability of hollow-core PCFs (HC-PCF) [1, 3] to host such gas-phase materials offers a promising alternative for such applications. For example, soliton compression and third-harmonic generation was demonstrated in HC-PCF filled with xenon [4] or argon [5, 6] at pressures up to 10 atm. Because the majority of the guided light in an HC-PCF resides in the hollow core, the Raman contribution from the surrounding silica cladding is greatly lessened. A disadvantage, though, is that the low density of the typically used noble gas compared to that of the silica results in a filling medium with nonlinear contributions from the core that are well below that of a solid fused-silica-core fiber.

We propose that using supercritical monoatomic fluids as a filling medium can bring the nonlinearity up to the same order of magnitude as that of fused silica, while still maintaining low amounts of Raman scattering [7,8]. Furthermore, the large controllable density range associated with the supercritical fluid leads to a highly variable refractive index, allowing the guidance and dispersion properties of the system to be tailored to particular light sources or experimental requirements.

Xenon makes a unique choice for observing such phenomena in HC-PCF. First, of the practical noble gases it has the highest nonlinear refractive index for a given density ($n_2 \approx 6.39 \times 10^{-19}$ cm$^2$/W at STP and 800 nm) [9], although still typically small compared to that of fused silica ($n_2 \approx 2.9 \times 10^{-16}$ cm$^2$/W) [10]. Second, its easily accessible critical point at 57.6 atm and 16.6 °C makes it convenient to create supercritical xenon with a density several hundred times that at STP while working at room temperature, bringing its nonlinearity up near that of fused silica. This, along with its negligible Raman scattering, makes supercritical xenon potentially a near-ideal platform for nonlinear optics and quantum optics.

In this paper we demonstrate ultra-high-density filling of a hollow-core photonic bandgap (HC-PBG) fiber with xenon. We show that the supercritical transition allows sensitive pressure tuning of the guidance spectral window, and give estimates for the nonlinearity and Raman processes in the system.

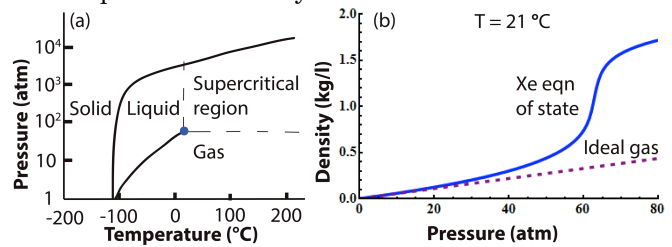

Fig 1. (color online) (a) Sketch of phase diagram of Xe, with the critical point highlighted in blue. (b) Density vs pressure for xenon and an ideal gas at temperature 21 °C.

The supercritical phase occurs in that region of a phase diagram above the critical point, Fig. 1(a), where gas and liquid become indistinguishable. Near the critical point, attractive interatomic forces cause a high fluid compressibility, leading to rapidly increasing density versus pressure, illustrated for Xe in Fig. 1(b), which shows that the density at 21 °C and 80 atm is 4.1 times higher than for an ideal gas [11]. Assuming a linear relationship with density, we calculate the nonlinearity of Xe to be $n_2 \approx 2.0 \times 10^{-16}$ cm²/W, or 69% of that of fused silica [12].

In HC-PBG fiber the cladding, consisting of high-refractive-index silica struts and low-index gas-filled holes, forms a 2-D photonic crystal surrounding the hollow gas-filled core (inset, Fig. 2). For certain wavelengths, there exists an out-of-plane bandgap [5]. When light with these wavelengths is coupled into the core, the bandgap prevents leakage through the cladding, and the light is transmitted with low loss.

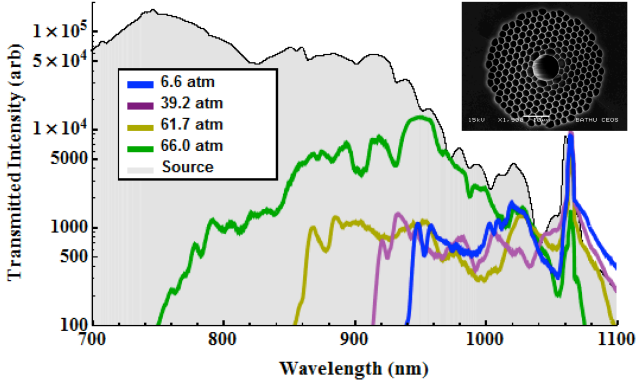

Fig. 2. (Color online) Four transmission spectra from the same fiber under different pressures of Xe, demonstrating the shift of the guidance window. The source spectrum is shown in the shaded background, attenuated to fit the relative intensity scale of the transmission spectra.

The density increase that occurs as Xe gas transitions from a normal to supercritical fluid leads to large changes in the guidance properties of the HC-PBG fiber. To model this we use a simple scaling law approximately describing how features of the HC-PBG fiber change for different values of the hole refractive index, obtained by Birks et al, for the case of a scalar wave equation [13]. This has been used successfully to predict how the bandgap of HC-PBG fiber shifts to new wavelengths when changing the filling-fluid refractive index [14]. Given that the bandgap edge is located at free-space wavelength $\lambda_{BGE}$ if the holes are filled with a medium having refractive index $n_0$, when the filling index is changed to $\overline{n}$ the edge is predicted to shift to wavelength

$$\overline{\lambda}_{BGE} = \lambda_{BGE} \sqrt{\frac{1 - \overline{n}^2 / n_S^2}{1 - n_0^2 / n_S^2}} \qquad (1)$$

where $n_S$ is the index of silica. The three index values in this relation depend on wavelength, through Sellmeier formulas, and the index of the Xe in the holes increases with density according to the Lorentz-Lorenz equation. Solving these self consistently, we predict and verify experimentally that the band edge depends very sensitively on Xe pressure when crossing into the supercritical regime at room temperature.

To demonstrate experimentally the shifting of the guidance window in the Xe supercritical regime, we coupled a fiber-laser-based white-light source into a Xe-filled HC-PBG fiber, with core diameter of 10 microns and guidance window centered at 1064 nm when filled with standard atmosphere. The open fiber ends were contained in high-pressure cells with windows, which were connected to a Xe gas system to allow control of the common pressure within the core and cladding holes of the fiber. The spectrum of the light transmitted by the fiber was spatially filtered, to ensure detection of light only emerging from the core, and measured with a spectrometer, with results shown in Fig. 2.

As the pressure of Xe was varied, the (unnormalized) transmission spectrum demarked the guidance window of the fiber, allowing us to observe the transition to supercritical fluid.

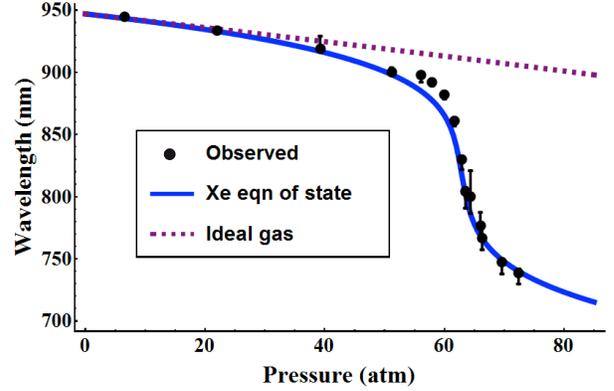

Fig. 3. (Color online) Theoretical predictions and experimental results for the shift in the guidance window blue edge of a HC-PBG fiber filled with Xe, showing the transition to supercritical fluid. The prediction of the ideal gas law is shown for comparison. Error bars indicate uncertainty in determining the edge of the guidance window.

The measured guidance window blue edge vs pressure is plotted in Fig. 3, where we compare it to the theoretical prediction of Eq.(1), for temperature of 21 °C (as measured to within 1 °C). The results show a shift in the guidance window by over 200 nm. Also plotted is the predicted edge shift for an ideal gas under similar pressures. Clear evidence of the supercritical transition is seen, in good agreement with the simple scaling theory using no free parameters.

To illustrate the versatility and practicality of the band-shifting technique, we also show in Fig. 4 the results of filling a HC-PBG fiber whose guidance window centers at 1550 nm at standard atmosphere. The band edges are found to shift in accordance with Eq.(1).

Given that the nonlinearity of the Xe in the core is calculated to be comparable to that of fused silica, and that atomic Xe has no Raman effect, any residual Raman effect will arise from the small intensity of light in the portion of the guided mode that overlaps the silica struts in the cladding. Experimental studies of vacuum-filled

HC-PBG fiber similar to that used here found the fraction of power interacting with the silica is in the range 0.0001 - 0.001 [15]. The value is sensitive to the particular fiber design, but remains negligible. An application of the scaling law in [13] predicts that for HC-PBG fiber the transverse wave number, and thus the fraction of power interacting with the silica, is relatively insensitive to the filling-fluid refractive index. We can thus confidently predict that Raman effects will be far less in this system than in fused-silica fibers.

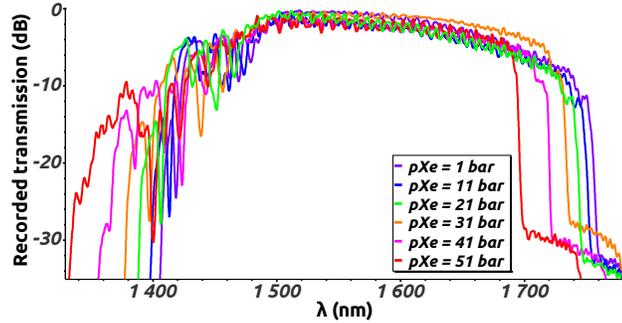

Fig. 4. (Color online) Six transmission spectra from a 1550 nm HC-PBG fiber under different pressures of Xe.

For many nonlinear optical processes such as photon pair generation and soliton propagation, it is necessary to work near a wavelength where the group-velocity dispersion (GVD) is zero. Equation (1), along with the Sellmeier and Lorentz-Lorenz equations, also predicts how the zero-GVD wavelength changes with filling index. It is known that in air-filled HC-PBG fiber the dispersion exhibits a zero-GVD wavelength located within the transmission window [5]. The scaling law predicts that the zero-dispersion wavelength will remain in its relative position within the guidance window as the window shifts to lower wavelength as done in this study. This allows the dispersion of the HC-PBG system to be controlled with small changes to pressure when near the Xe phase transition. (This differs from the case of gas-filled Kagome' fiber, which has a zero-dispersion wavelength that shifts to higher wavelengths with increased pressure [6].)

In conclusion, the creation of supercritical xenon within an HC-PBG fiber was experimentally demonstrated. By using supercritical Xe, the optical nonlinearity of the system at easily achievable pressures and temperatures is predicted to approach that of fused silica while maintaining a negligible amount of Raman scattering. In addition, the large variation of the xenon density when going through the phase transition can be used to tailor the guidance window and dispersion of the system via small changes of pressure. By operating in this range, the system can be tuned to a particular light source, allowing transmission of light originally not guided by the fiber, and shifting the zero-dispersion point to a desired wavelength. It is apparent that the use of supercritical fluids in HC-PBG and other hollow-core fibers has great potential for creating nearly ideal highly nonlinear optical systems.

We thank John Hardwick for suggesting Xe as a supercritical fluid. We acknowledge support from NSF Physics (USA) and la région limousine (France).